\documentclass[11 pt,a4paper]{article}
\usepackage[utf8]{inputenc}
\usepackage{amsmath}
\usepackage{amsfonts}
\usepackage{amssymb}
\usepackage{pgfplots}
\usepackage{tikz}
\usepackage{subcaption}
\usepackage{amsmath}
\usepgfplotslibrary{fillbetween}
\usetikzlibrary{patterns}
\usepackage{indentfirst}
\usepackage{amsmath}
\usepackage{booktabs}
\usepackage{pgfplotstable}

\author{Dr. Jacques Balayla MD, MPH, CIP, FRCSC\footnote{To whom correspondence should be addressed: Dr. Jacques Balayla MD, MPH, CIP, FRCSC. Quilligan Scholar. e-mail: jacques.balayla@mail.mcgill.ca. Department of Obstetrics and Gynaecology. McGill University, Montreal, Quebec, Canada}}

\title{\LARGE Bayesian Updating and Sequential Testing:  Overcoming Inferential Limitations of Screening Tests}

\date{}

\begin{document}
\maketitle  
\begin{abstract}

Bayes' Theorem confers inherent limitations on the accuracy of screening tests as a function of disease prevalence. We have shown in previous work that a testing system can tolerate significant drops in prevalence, up until a certain well-defined point known as the \textit{prevalence threshold}, below which the reliability of a positive screening test drops precipitously. Herein, we establish a mathematical model to determine whether sequential testing with a single test overcomes the aforementioned Bayesian limitations and thus improves the reliability of screening tests. We show that for a desired positive predictive value of $\rho$ approaching $k$, the number of positive test iterations needed is:
\begin{large}
\begin{equation}
n_i =\lim_{\rho \to k}\left\lceil\frac{ln\left[\frac{\rho(\phi-1)}{\phi(\rho-1)}\right]}{ln\left[\frac{a}{1-b}\right]}\right\rceil 
\end{equation}
\end{large}
\

where $n_i$ = number of testing iterations necessary to achieve $\rho$, the desired positive predictive value, ln = the natural logarithm, a = sensitivity, b = specificity, $\phi$ = disease prevalence/pre-test probability and $k$ = constant. Based on the aforementioned derivation, we provide reference tables for the number of test iterations needed to obtain a $\rho(\phi)$ of 50, 75, 95 and 99$\%$ as a function of various levels of sensitivity, specificity and disease prevalence/pre-test probability. 

\end{abstract} 
\newpage

\section{Bayes’ Theorem}
Bayes’ Theorem describes the probability of an event, based on prior knowledge of conditions that are related to the event \cite{hall1967clinical}. As a principle, it follows simply from the axioms of conditional probability \cite{rouder2018teaching}. Mathematically speaking, the equation translates to the conditional probability of an event A given the presence of an event or state B. Indeed, as per Bayes' Theorem, the above relationship is equal to the probability of event B given event A, multiplied by the ratio of independent probabilities of event A to event B \cite{rouder2018teaching}. Simply stated, the equation is written as follows:

\begin{large}
\begin{equation}
P(A|B) = \frac{P(B|A) P(A)}{P(B)}
\end{equation}
\end{large}

Where A, B = events, P(A) and P(B) are the independent probabilities of A and B, $P(A|B)$ = probability of A given B is true and $P(B|A)$ = probability of B given A is true.
\subsection{Proof of Bayes' Theorem and its Relationship to $\rho(\phi)$}

Let us denote to independent events, A and B. The probability of events A and B both occurring is denoted axiomatically as $P(A\cap B)$, and it equals to the conditional probability of A, P(A), times the probability of B given that A has occurred, $P(B|A)$.

\begin{large}
\begin{equation}
P(A \cap B) = P(A)P(B|A)
\end{equation}
\end{large} 
Likewise, since we have pre-conditionally established that both events are occurring, the index event order is commutative and could be switched to obtain:
\begin{large}
\begin{equation}
P(A \cap B) = P(B)P(A|B)
\end{equation}
\end{large} 
Equating the terms, we obtain the formal Bayes' theorem as follows \cite{dezert2018total}:
\begin{large}
\begin{equation}
P(A|B) = \frac{P(B|A) P(A)}{P(B)}
\end{equation}
\end{large} 

If we use T +/- as either a positive or negative test, and denote D +/- as the presence (+) or absence (-) of disease then we can use Bayes' theorem to calculate the positive predictive value (PPV) of a screening test by asking the following: given a positive screening test result, what is the probability that such individual does in fact have the disease in question? In other words, what is the probability that a positive test is a true positive? \cite{lutgendorf201699}. 

\begin{large}
\begin{equation}
P(D+|T+) = \frac{P(T+|D+) P(D+)}{P(T+)}
\end{equation}
\end{large} 

Since the denominator in equation (6) represents the probability of having a positive test regardless of context, then it follows logically that this variable should equal to the sum of true positives and false positives.
\\
\

Otherwise stated:
\begin{large}
\begin{equation}
P(D+|T+)= \frac{P(T+|D+)P(D+)}{P(T+|D+)P(D+)+P(T+|D-)P(D-)}
\end{equation}
\end{large} 

Furthermore, given that 1) the probability of having a positive test in an individual with the disease is a test's sensitivity, and 2) the probability of being disease-free is equal to the complement of the prevalence, and 3) the false positive rate is equal to the complement of the specificity (true negative rate), Bayes' theorem provides a formal way to obtain the PPV, $\rho(\phi)$, as a function of the prevalence $\phi$, as follows \cite{simon1990sensitivity}:

\begin{large}
\begin{equation}
\rho(\phi)= \frac{a\phi}{ a\phi+(1-b)(1-\phi)} =\frac{a\phi}{a\phi+ b\phi - b - \phi +1}
\end{equation}
\end{large} 
\\
\
where $\rho(\phi)$ = PPV, a = sensitivity, b = specificity and $\phi$ = prevalence.
\\
\

We have thus shown that the PPV, $\rho(\phi)$, is a function of prevalence, $\phi$. As the prevalence increases, the $\rho(\phi)$ also increases and vice-versa. By the above equation, we obtain:

\begin{large}
\begin{equation}
\lim_{\phi \to 0} \rho(\phi) = 0
\end{equation}
\end{large}
\begin{large}
\begin{equation}
\lim_{\phi \to 1} \rho(\phi) = 1
\end{equation}
\end{large}

These limits denote the extremes of domain of the function $\rho(\phi)$, notably [0,0], and [1,1]. Conversely, using the same derivation technique, the negative predictive value, $\sigma(\phi)$ can be denoted as \cite{simon1990sensitivity}:
\begin{large}
\begin{equation}
\sigma(\phi) = \frac{b(1-\phi)}{(1-a)\phi+b(1-\phi)} 
\end{equation}
\end{large}
\\
\
The extreme limits of the domain of this function include [0,1] and [1,0].
\newpage
\section{Prevalence Threshold}
Let us define $\varepsilon = a +b$, the sum of a tests' sensitivity and specificity. Likewise, let us define the point of local extrema [$\phi_e,\rho_e$], which denotes the point of maximum curvature of the function $\rho(\phi)$ throughout the prevalence spectrum as per equation (8). Practically speaking, $\phi_e$ tells us where the sharpest turn, or change, in PPV as a function of prevalence occurs. In cases of $\varepsilon <$ 1 the sharp increase occurs at higher prevalence levels with low PPV levels. Conversely, when $\varepsilon >$ 1, as in the example below, the sharp increase occurs at lower prevalence levels with high PPV levels \cite{balayla2020prevalence}.
\\
\

\begin{tikzpicture}[trim left=-4.3cm]
 
	\begin{axis}[
    axis lines = left,
    xlabel = $\phi$,
	ylabel = {$\rho(\phi)$},    
     ymin=0, ymax=1,
    legend pos = south east,
     ymajorgrids=true,
    xmajorgrids=true,
    width=5cm,
    height=5cm,
     ]

\addplot [	
	domain= 0:1,
	color= blue,
	]
     {(0.98*x)/((0.98*x+(1-0.97)*(1-x))}; 
	\node[above,blue] at (680,20) {$\phi_e$ = 0.15};
	\addlegendentry{$\varepsilon = 1.95$}
\end{axis}
\end{tikzpicture}
\begin{center}
\begin{small}
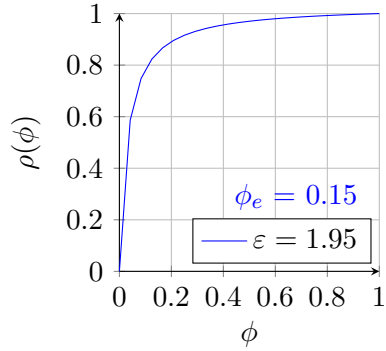
Figure 1. Sample test with $\varepsilon$ $>$ 1, where a = 0.98 and b = 0.97
\end{small}
\end{center}
\subsection{How is $\phi_e$ determined?}
Using differential calculus, as showcased in previous work by this author \cite{balayla2020prevalence}, the formula for $\phi_e$ is defined as follows:
\\
\

\begin{large}
\begin{equation}
\phi_e = \frac{\sqrt{a\left(-b+1\right)}+b-1}{(a+b-1)}=\frac{\sqrt{a\left(-b+1\right)}+b-1}{(\varepsilon-1)}
\end{equation}
\end{large}

This is the value of prevalence as function of sensitivity and specificity where the point of local extrema $\phi_e$ is found. We denote this value of $\phi$ as the \textit{prevalence threshold}. By plugging $\phi_e$ into equation (3) we obtain its corresponding $\rho_e$ value. Note the inverse relationship between $\phi_e$ and $\varepsilon$.

\begin{large}
\begin{equation}
\phi_e \sim \frac{1}{\varepsilon}
\end{equation}
\end{large}

From the above relationship it follows that even with values of sensitivity and specificity whose sum nearly approaches its maximum value of 2, there is a precipitous drop in PPV at low prevalence levels where the test's $\rho$ is significantly hindered.
\section{Sequential Testing}

Based on the aforementioned considerations, a problem arises. Since the vast majority of medical conditions and disorders amenable to screening have prevalences that are low in the general population, we deduce that a significant proportion of positive screening tests conducted in modern practice are false positives, which can bring about significant adverse administrative, social, health and psychological consequences. As such, this insurmountable fact about the nature of screening begs the question - is there anything to be done to reduce the number of incorrect diagnoses that arise given the aforementioned limitation? \cite{moons1997limitations}. Intuitively, as per equation (8), the development of novel screening tests with parameters such that $\varepsilon\sim2$ would reduce the influence of prevalence in the equation \cite{balayla2020prevalence}. But such endeavour is costly and most often unattainable in the short term. Given human error, variations in patient status/characteristics, sampling error and technological limitations, the most intuitive method to ensure a correct diagnosis is made on a patient is that of sequential, or repetitive, testing \cite{woloshin2020false}. This phenomenon is technically known as $Bayesian$ $updating$. While this is a general term that is used when any new information is added onto a system which was previously analysed, it too applies when the same test is run serially to improve its detection rate.
\subsection{Conditional Probabilities}
Conditional probabilities relate the likelihood of an occurrence given that another related event has already taken place. That initial condition is termed \textit{prior probability} or in certain circumstances the \textit{pre-test probability}. When we account for those prior probabilities, and analyse a screening test in that context, we obtain \textit{posterior probabilities}. In general, with sequential Bayesian estimation, one can use the previous posterior as the current prior probability. As such, in the case of sequential testing where D represents the presence of disease, T represents one initial positive test and TT represents two consequent positive tests, Bayes' theorem takes on the form:
\\
\
\begin{large}
\begin{equation}
P(D|T) = \frac{P(T|D)P(D)}{P(T)} \Rightarrow P(D|TT) = \frac{P(TT|D)P(D)}{P(TT)}
\end{equation}
\end{large} 
\newpage
\section{General Derivation}
The expression of equation (2) in generalized terms is the following:
\begin{large}
\begin{equation}
P(D|T) = \frac{P(T|D) P(D)}{ P(T|D) P(D) + P(T|\neg D) P(\neg D)}
\end{equation}
\end{large} 
where,
\begin{itemize}
	\item P(D) is the prior probability, or the initial degree of belief in D
	\item P($\neg D$) is the corresponding initial degree of belief  in  'not-D', where P($\neg D$) = 1-P(D)
	\item P(T$|$D) is the conditional probability or likelihood of T given that  proposition D is true
\end{itemize}
\subsection{Bayesian updating formulation}
Let $T_1,T_2,...,T_n$ denote n $independently$ conducted tests.
\begin{align*}
P(T_1T_2...T_n) &= P(T_1T_2...T_n|D)P(D) + P(T_1T_2...T_n|\neg D)P(\neg D)\\
&= P(T|D)^nP(D) + P(T|\neg D)^nP(\neg D)\\
&= P(T|D)^nP(D) + P(T|\neg D)^n(1-P(D))
\end{align*}

Then, we can find our expression for $P(D|T_1,T_2,...,T_n$).

\begin{align*}
P(D|T_1T_2...T_n) &= \dfrac{P(T_1T_2...T_n)|D)P(D)}{P(T_1T_2...T_n)}\\
&= \dfrac{P(T|D)^nP(D)}{P(T|D)^nP(D) + P(T|\neg D)^n(1-P(D))}
\end{align*}
It thus follows that as $n\rightarrow$ $\infty$, at some iteration $n_x$ the above equation converges as a function of $P(T|D)$:
\begin{align*}
\lim_{n\to\infty} \dfrac{P(T|D)^nP(D)}{P(T|D)^nP(D) + P(T|\neg D)^n(1-P(D))}=
\begin{cases} 
      1 & \text{if } P(T|D) > 0.5 \\
      P(D) & \text{if } P(T|D) = 0.5 \\
      0 & \text{if } P(T|D) < 0.5 
\end{cases}
\end{align*}
In terms of screening parameters, the above equation therefore becomes:
\begin{large}
\begin{equation}
\rho(\phi)= \frac{a^n\phi}{ a^n\phi+(1-b)^n(1-\phi)} = \frac{a^n\phi}{a^n\phi+(1-b)^n -\phi(1-b)^n} 
\end{equation}
\end{large}
\\
\
where $n$ is the number of test iterations.
\\
\
\newpage
To determine the number of tests needed to obtain a desired predictive value, we need to first isolate n by re-arranging equation (16) as follows:

\begin{large}
\begin{equation}
\rho(\phi)a^n\phi+\rho(\phi)(1-b)^n(1-\phi)={a^n\phi}
\end{equation}
\end{large}
Re-arranging the terms:
\begin{large}
\begin{equation}
\rho(\phi)a^n\phi-a^n\phi = -\rho(\phi)(1-b)^n(1-\phi)
\end{equation}
\end{large}
Factoring out the sensitivity $a$:
\begin{large}
\begin{equation}
a^n\phi[\rho(\phi)-1] = -\rho(\phi)(1-b)^n(1-\phi)
\end{equation}
\end{large}
By the fraction rule of exponents:
\begin{large}
\begin{equation}
\frac{a^n}{(1-b)^n} = \frac{-\rho(\phi)(1-\phi)}{\phi[\rho(\phi)-1]} = \left[\frac{a}{1-b}\right]^n
\end{equation}
\end{large}
Applying the natural logarithm ($ln$) to both sides:
\begin{large}
\begin{equation}
ln\left[\frac{-\rho(\phi)(1-\phi)}{\phi[\rho(\phi)-1]}\right] = ln\left[\frac{a}{1-b}\right]^n 
\end{equation}
\end{large}
Via the power rule, we obtain:
\begin{large}
\begin{equation}
ln\left[\frac{-\rho(\phi)(1-\phi)}{\phi[\rho(\phi)-1]}\right] = ln\left[\frac{a}{1-b}\right]n 
\end{equation}
\end{large}
From the above relationship, we can isolate n:
\begin{large}
\begin{equation}
n = \frac{ln\left[\frac{-\rho(\phi)(1-\phi)}{\phi[\rho(\phi)-1]}\right]}{ln\left[\frac{a}{1-b}\right]}
\end{equation}
\end{large}
Finally, simplifying the expression:
\begin{large}
\begin{equation}
n = \frac{ln\left[\frac{\phi\rho(\phi)-\rho(\phi)}{\phi\rho(\phi)-\phi}\right]}{ln\left[\frac{a}{1-b}\right]}
\end{equation}
\end{large}
From this expression we can calculate the limit as $\rho(\phi)$ goes to 1, the ultimate predictive value:
\begin{large}
\begin{equation}
n = \lim_{\rho(\phi) \to 1}\frac{ln\left[\frac{\phi\rho(\phi)-\rho(\phi)}{\phi\rho(\phi)-\phi}\right]}{ln\left[\frac{a}{1-b}\right]} 
\end{equation}
\end{large}
However, the $\lim_{\rho(\phi) \to 1} n$ does not exist,  since ln($\phi$-1/0) is undefined. In clinical terms, this translates to the fact that in all but one special case where disease prevalence $\phi$ is 1, no test can have a perfect positive predictive value.
\newpage
To overcome this limitation, we render the generalized form of the above equation, and we denote $\rho(\phi)$ as $\rho$ to obtain:
\begin{large}
\begin{equation}
n_i =\lim_{\rho \to k}\frac{ln\left[\frac{\rho(\phi-1)}{\phi(\rho-1)}\right]}{ln\left[\frac{a}{1-b}\right]} 
\end{equation}
\end{large}
\

where $\rho$ = desired positive predictive value to achieve, $n_i$ = number of testing iterations necessary, a = sensitivity, b = specificity, $\phi$ = disease prevalence and $k$ = constant. 
\section{Positive Likelihood Ratio - LR+}
From equation (26) we observe that the number of serial tests $n$ needed to attain a given PPV value is inversely proportional to $ln\left[\frac{a}{1-b}\right]$. The latter expression in brackets represents what is known as the positive likelihood ratio (+LR) \cite{mcgee2002simplifying}. A likelihood ratio (LR) for a dichotomous test is defined as the likelihood of a test result in patients with the
disease divided by the likelihood of the test result in patients without the disease. Otherwise stated, the positive likelihood ratio (+LR) gives the change in the odds of having a diagnosis in patients with a positive test. For example, a LR+ close to 1 means that the test result does not change the likelihood of disease or the outcome of interest appreciably. The more the likelihood ratio for a positive test (LR+) is
greater than 1, the more likely the disease or outcome \cite{mcgee2002simplifying}. It would thus follow that the greater the likelihood ratio of a test the lower number of sequential tests needed to achieve a particular PPV. 
\section{Properties of sequential testing}
Since the natural logarithmic function is continuous and increasing throughout its domain (0,$\infty_+$], it follows that as $ln\left[\frac{a}{1-b}\right]$ increases, the number of test iterations $n$ needed to achieve a desired positive predictive value decreases as per equation (26). Tables 1-4 provide different reference values of $n$ as a function of the prevalence $\phi$ and the sensitivity and specificity for a $\rho$ of 99, 95, 75 and 50 $\%$, respectively. Figure 3 provides a graphic representation of the $n_i$, which given its geometric shape we define as the $tablecloth$ function. The aforementioned relationship holds for a number of identical sequential tests that are positive until the $n_i$ iteration reaches the desired positive predictive value. For severe conditions whose treatment is rather innocuous but whose potential consequences are severe, a lower threshold to initiate treatment might be acceptable. Conversely, a condition whose consequences are less severe but whose treatment may lead to significant morbidity might benefit from a higher degree of diagnostic certainty prior to initiating therapy or proceeding to an invasive diagnostic test. Given the extremes of the domains of each predictive function as per equations (8) and (11), and the fact that most conditions have a prevalence well below 20$\%$ then it follows that if prior to reaching the desired positive predictive value, a negative test result is obtained, the individual is more likely to be disease-free, since $\sigma(\phi)$ $>>$ $\rho(\phi)$ at a low prevalence level of disease. It is critical to bear in mind that testing might be done in a representative sample of a population to estimate the rate of asymptomatic carriage; in this case the prevalence is meaningful.  But testing is generally done in subjects in whom a condition is suspected, either because they have a known exposure or because they have various levels of symptomatology. In such cases the population prevalence is irrelevant, and it would be more appropriate to refer to prior probability instead.
\
\begin{center}
\begin{tikzpicture}
	\begin{axis}[
    axis lines = left,
    xlabel = $\phi$,
	ylabel = {$\rho(\phi)$,$\sigma(\phi)$},    
     ymin=0, ymax=1,
    legend pos = outer north east,
     ymajorgrids=true,
     xmajorgrids=true,
    grid style=dashed,
    width=7cm,
    height=5cm,
     ]
	\addplot [
	domain= 0:1,
	color= blue,
	]
	{(0.80*x)/((0.80*x+(1-0.85)*(1-x))};

\addplot [	
	domain= 0:1,
	color= red,
	]
	{0.85*(1-x)/((1-0.8)*x+0.85*(1-x))};
	
\addplot+ [
dashed,
domain= 0:1,	
color = black,
mark size = 0pt
 ]
	{1};
	\addlegendentry{$\rho(\phi)$}
	\addlegendentry{$\sigma(\phi)$}
	\addlegendentry{$\rho,\sigma=1$}
	\end{axis}
\end{tikzpicture}
\end{center}
\begin{center}
\begin{small}
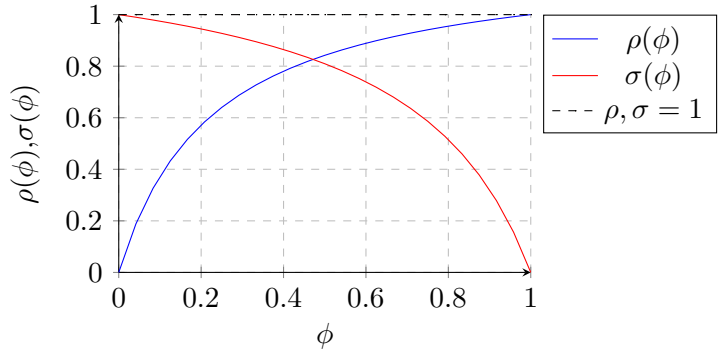
\textbf{Figure 2}. Overlapping positive (blue) and negative (red) predictive value curves.
\end{small}
\end{center}
\subsection{Clinical Implications of $n_i$}
From the formula in (26), we learn that the number of iterations is inversely proportional to the ratio of sensitivity over the complement of the specificity - which represents the +LR \cite{mcgee2002simplifying}.

\begin{large}
\begin{equation}
n_i\sim{\frac{1}{ln\left[\frac{a}{1-b}\right]}} 
\end{equation}
\end{large}

However, the denominator of this equation is itself the natural logarithm of a fraction. It follows that for certain values of sensitivity a and specificity b, the ratio of $[\frac{a}{1-b}]$ is $<$ 1. Since the natural logarithm of $x$ follows the following range properties:

\begin{large}
\begin{equation}
\ln(x) =
\begin{cases} 
	  \in \mathbb{C} & \text{if } x \leq 0  \\
	  undefined & \text{if } x = 0  \\       
      < 0 & \text{if } 0 < x < 1 \\
      \geq 0 & \text{if } x \geq 1\\
\end{cases}
\end{equation}
\end{large}
We deduce that for values of a and b such that:
\\
\
\begin{large}
\begin{equation}
a<1-b \Leftrightarrow a + b < 1 
\end{equation}
\end{large}
the denominator of the $n_i$ function will be negative and so will thus be $n_i$.
\\
\

Though it is unlikely that a test whose sensitivity and specificity add to less than one would be often used clinically \cite{grimes2002uses}, this idea does lead to a fundamental understanding about the $n_i$ equation. What does it mean to have a negative number of tests needed to achieve a given $\rho(\phi)$? Clinically it bears no meaning, since one would, by definition, need at least a single test to have a positive result. It thus follows that for the above equation to be of clinical use, we need to take its ceiling function \cite{weisstein2002ceiling}, such that $\lceil x \rceil$ is the unique integer satisfying $\lceil x \rceil$ - 1 $<$ x $<$ $\lceil x \rceil$:

\begin{large}
\begin{equation}
n_i =\lim_{\rho \to k}\left\lceil\frac{ln\left[\frac{\rho(\phi-1)}{\phi(\rho-1)}\right]}{ln\left[\frac{a}{1-b}\right]}\right\rceil 
\end{equation}
\end{large}

In practical terms, the ceiling function assigns the nearest positive integer to a number \cite{weisstein2002ceiling}. For the case of screening tests, it implies that a whole rather than a decimal number of tests (rounded to the nearest, higher, positive integer) ought to be performed.
\section{Independence of Serial Testing}
From the concepts described in this work, one might easily suggest that simply repeating the same screening test multiple times increases confidence that a positive result is a true positive. Setting aside the administrative and feasibility concerns, while such an interpretation is theoretically correct, the reality ought to be more nuanced, as there are confounding factors that might make the same result recur upon serial testing on the same patient. Indeed, repeating the same test under the same conditions, in a similar time-frame, perhaps even by the same interpreter/provider may not constitute a true independent observation \cite{balayla2020derivation}. As such, the primary use of the tables and notions herein described ought to be to contextualize the screening result and broaden the clinical judgement of the provider with regards to the reliability of the screening process. A more natural and reliable method to enhance the positive predictive value would be, when available, to use a different test with different parameters altogether after an initial positive result is obtained  \cite{balayla2020derivation}.
\section{Conclusion}
In this manuscript, we describe a mathematical model to determine whether sequential testing with a single test overcomes the Bayesian limitations of screening and thus improves the reliability of screening tests. We show that for a desired positive predictive value of $\rho$ that approaches $k$, the number of positive test iterations $n_i$ is inversely proportional to the natural logarithm of the positive likelihood ratio (LR+). This clinical utility of this equation is best observed in conditions with low pre-test probability where single tests are insufficient to achieve clinically significant predictive values and likewise, in clinical scenarios with a high pre-test probability where confirmation of disease status is critical. When independent observations are difficult to obtain, serial testing with a different test will likewise enhance the positive predictive value.
\newpage
\begin{tikzpicture}[scale=1.50]
\begin{axis}[grid=major,view={500}{10}, colormap/cool, minor tick num=4, tick label style={font=\scriptsize, /pgf/number format/fixed},label style={font=\footnotesize},
xtick={0,1,2,3,4,5},
ytick={0,.05,.1,.15,.2},
ztick={0,2,4,6,8,10,12,14,16,18},
 xlabel = $\ln(\frac{a}{1-b})$,
  ylabel = Prevalence ($\phi$),
  zlabel = {$n_i$},
  title = \Large $n_i$ iteration plot]
  xmin = 0,   
  xmax = 0.2,
  ymin = 0,   
  ymax = 5.0,
  zmin = 0,   
  zmax = 20,

  \addplot3+ [surf, mark options={scale=.1}] table {table.dat};
\end{axis}
\end{tikzpicture}
\\
\

\begin{center}
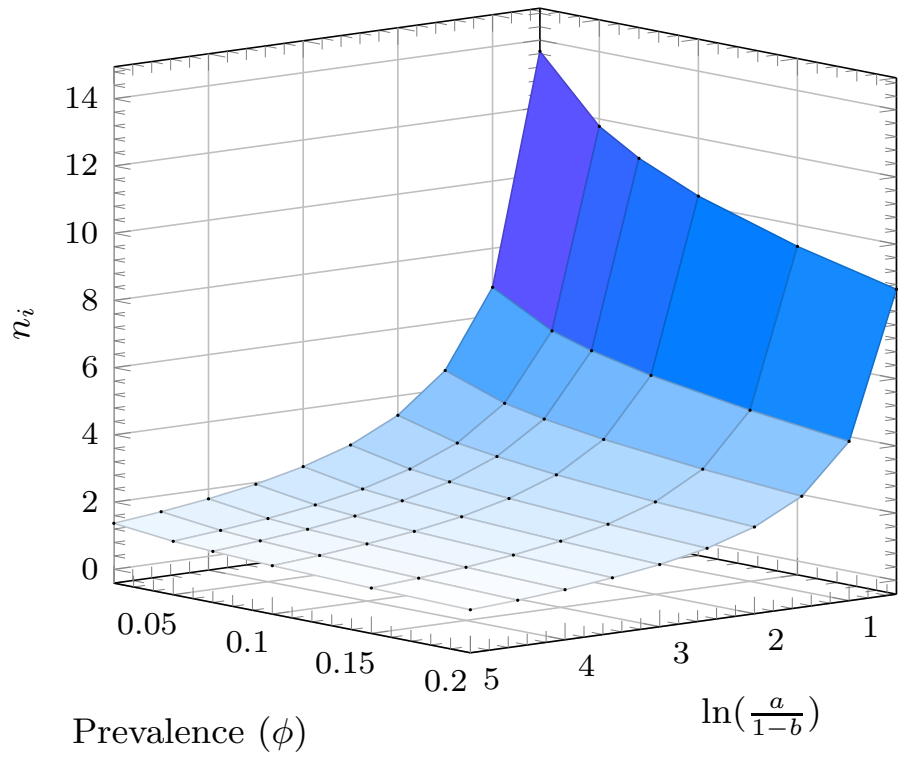
Figure 3. $n_i$ iteration plot as a function of sensitivity a, specificity b, and disease prevalence $\phi$ for a positive predictive value of 95$\%$.
\end{center}

\newpage
\section{Addendum - Reference Tables}
\begin{table}[h!]
\centering
\begin{tabular}{|c|c|c|c|c|c|c|}
\hline
\multicolumn{1}{|l|}{} & \multicolumn{6}{c|}{\textbf{Prevalence $(\phi)$}}                                                    \\ \hline
\textbf{$\ln(\frac{a}{1-b})$}     & \textbf{0.02} & \textbf{0.05} & \textbf{0.07} & \textbf{0.1} & \textbf{0.15} & \textbf{0.2} \\ \hline
\textbf{0.50}          & 16.97         & 15.08         & 14.36         & 13.58        & 12.66         & 11.96        \\ \hline
\textbf{1.00}          & 8.49          & 7.54          & 7.18          & 6.79         & 6.33          & 5.98         \\ \hline
\textbf{1.50}          & 5.66          & 5.03          & 4.79          & 4.53         & 4.22          & 3.99         \\ \hline
\textbf{2.00}          & 4.24          & 3.77          & 3.59          & 3.40         & 3.16          & 2.99         \\ \hline
\textbf{2.50}          & 3.39          & 3.02          & 2.87          & 2.72         & 2.53          & 2.39         \\ \hline
\textbf{3.00}          & 2.83          & 2.51          & 2.39          & 2.26         & 2.11          & 1.99         \\ \hline
\textbf{3.50}          & 2.42          & 2.15          & 2.05          & 1.94         & 1.81          & 1.71         \\ \hline
\textbf{4.00}          & 2.12          & 1.88          & 1.80          & 1.70         & 1.58          & 1.50         \\ \hline
\textbf{4.50}          & 1.89          & 1.68          & 1.60          & 1.51         & 1.41          & 1.33         \\ \hline
\textbf{5.00}          & 1.70          & 1.51          & 1.44          & 1.36         & 1.27          & 1.20         \\ \hline
\end{tabular}
\caption{Reference table for the number of test iterations to obtain a $\rho(\phi)$ of 99$\%$ as a function of sensitivity a, specificity b and disease prevalence $\phi$. To enhance the predictive value and perform a whole number of tests, round up to the nearest integer using the ceiling function $\lceil x \rceil$.}
\label{tab:my-table}
\end{table}
\begin{table}[h!]
\centering
\begin{tabular}{|c|c|c|c|c|c|c|}
\hline
\multicolumn{1}{|l|}{} & \multicolumn{6}{c|}{\textbf{Prevalence $(\phi)$}}                                                    \\ \hline
\textbf{$\ln(\frac{a}{1-b})$}     & \textbf{0.02} & \textbf{0.05} & \textbf{0.07} & \textbf{0.1} & \textbf{0.15} & \textbf{0.2} \\ \hline
\textbf{0.50}          & 13.67         & 11.78         & 11.06         & 10.28        & 9.36          & 8.66         \\ \hline
\textbf{1.00}          & 6.84          & 5.89          & 5.53          & 5.14         & 4.68          & 4.33         \\ \hline
\textbf{1.50}          & 4.56          & 3.93          & 3.69          & 3.43         & 3.12          & 2.89         \\ \hline
\textbf{2.00}          & 3.42          & 2.94          & 2.77          & 2.57         & 2.34          & 2.17         \\ \hline
\textbf{2.50}          & 2.73          & 2.36          & 2.21          & 2.06         & 1.87          & 1.73         \\ \hline
\textbf{3.00}          & 2.28          & 1.96          & 1.84          & 1.71         & 1.56          & 1.44         \\ \hline
\textbf{3.50}          & 1.95          & 1.68          & 1.58          & 1.47         & 1.34          & 1.24         \\ \hline
\textbf{4.00}          & 1.71          & 1.47          & 1.38          & 1.29         & 1.17          & 1.08         \\ \hline
\textbf{4.50}          & 1.52          & 1.31          & 1.23          & 1.14         & 1.04          & 0.96         \\ \hline
\textbf{5.00}          & 1.37          & 1.18          & 1.11          & 1.03         & 0.94          & 0.87         \\ \hline
\end{tabular}
\caption{Reference table for the number of test iterations to obtain a $\rho(\phi)$ of 95$\%$ as a function of sensitivity a, specificity b and disease prevalence $\phi$. To enhance the predictive value and perform a whole number of tests, round up to the nearest integer using the ceiling function $\lceil x \rceil$.}
\label{tab:my-table2}
\end{table}
\newpage
\begin{table}[h!]
\centering
\begin{tabular}{|c|c|c|c|c|c|c|}
\hline
\multicolumn{1}{|l|}{} & \multicolumn{6}{c|}{\textbf{Prevalence $(\phi)$}}                                                    \\ \hline
\textbf{$\ln(\frac{a}{1-b})$}     & \textbf{0.02} & \textbf{0.05} & \textbf{0.07} & \textbf{0.1} & \textbf{0.15} & \textbf{0.2} \\ \hline
\textbf{0.50}          & 9.98          & 8.09          & 7.37          & 6.59         & 5.67          & 4.97         \\ \hline
\textbf{1.00}          & 4.99          & 4.04          & 3.69          & 3.30         & 2.83          & 2.48         \\ \hline
\textbf{1.50}          & 3.33          & 2.70          & 2.46          & 2.20         & 1.89          & 1.66         \\ \hline
\textbf{2.00}          & 2.50          & 2.02          & 1.84          & 1.65         & 1.42          & 1.24         \\ \hline
\textbf{2.50}          & 2.00          & 1.62          & 1.47          & 1.32         & 1.13          & 0.99         \\ \hline
\textbf{3.00}          & 1.66          & 1.35          & 1.23          & 1.10         & 0.94          & 0.83         \\ \hline
\textbf{3.50}          & 1.43          & 1.16          & 1.05          & 0.94         & 0.81          & 0.71         \\ \hline
\textbf{4.00}          & 1.25          & 1.01          & 0.92          & 0.82         & 0.71          & 0.62         \\ \hline
\textbf{4.50}          & 1.11          & 0.90          & 0.82          & 0.73         & 0.63          & 0.55         \\ \hline
\textbf{5.00}          & 1.00          & 0.81          & 0.74          & 0.66         & 0.57          & 0.50         \\ \hline
\end{tabular}
\caption{Reference table for the number of test iterations to obtain a $\rho(\phi)$ of 75$\%$ as a function of sensitivity a, specificity b and disease prevalence $\phi$. To enhance the predictive value and perform a whole number of tests, round up to the nearest integer using the ceiling function $\lceil x \rceil$.}
\label{tab:my-table3}
\end{table}
\begin{table}[h!]
\centering
\begin{tabular}{|c|c|c|c|c|c|c|}
\hline
\multicolumn{1}{|l|}{} & \multicolumn{6}{c|}{\textbf{Prevalence $(\phi)$}}                                                    \\ \hline
\textbf{$\ln(\frac{a}{1-b})$}     & \textbf{0.02} & \textbf{0.05} & \textbf{0.07} & \textbf{0.1} & \textbf{0.15} & \textbf{0.2} \\ \hline
\textbf{0.50}          & 7.78          & 5.89          & 5.17          & 4.39         & 3.47          & 2.77         \\ \hline
\textbf{1.00}          & 3.89          & 2.94          & 2.59          & 2.20         & 1.73          & 1.39         \\ \hline
\textbf{1.50}          & 2.59          & 1.96          & 1.72          & 1.46         & 1.16          & 0.92         \\ \hline
\textbf{2.00}          & 1.95          & 1.47          & 1.29          & 1.10         & 0.87          & 0.69         \\ \hline
\textbf{2.50}          & 1.56          & 1.18          & 1.03          & 0.88         & 0.69          & 0.55         \\ \hline
\textbf{3.00}          & 1.30          & 0.98          & 0.86          & 0.73         & 0.58          & 0.46         \\ \hline
\textbf{3.50}          & 1.11          & 0.84          & 0.74          & 0.63         & 0.50          & 0.40         \\ \hline
\textbf{4.00}          & 0.97          & 0.74          & 0.65          & 0.55         & 0.43          & 0.35         \\ \hline
\textbf{4.50}          & 0.86          & 0.65          & 0.57          & 0.49         & 0.39          & 0.31         \\ \hline
\textbf{5.00}          & 0.78          & 0.59          & 0.52          & 0.44         & 0.35          & 0.28         \\ \hline
\end{tabular}
\caption{Reference table for the number of test iterations to obtain a $\rho(\phi)$ of 50$\%$ as a function of sensitivity a, specificity b and disease prevalence $\phi$. To enhance the predictive value and perform a whole number of tests, round up to the nearest integer using the ceiling function $\lceil x \rceil$.}
\label{tab:my-table4}
\end{table}
\newpage
\bibliographystyle{unsrt}
\bibliography{bibliography.bib}

\begin{thebibliography}{10}

\bibitem{hall1967clinical}
GH~Hall.
\newblock The clinical application of bayes' theorem.
\newblock {\em The Lancet}, 290(7515):555--557, 1967.

\bibitem{rouder2018teaching}
Jeffrey~N Rouder and Richard~D Morey.
\newblock Teaching bayes’ theorem: Strength of evidence as predictive
  accuracy.
\newblock {\em The American Statistician}, 2018.

\bibitem{dezert2018total}
Jean Dezert, Albena Tchamova, and Deqiang Han.
\newblock Total belief theorem and generalized bayes' theorem.
\newblock In {\em 2018 21st International Conference on Information Fusion
  (FUSION)}, pages 1040--1047. IEEE, 2018.

\bibitem{lutgendorf201699}
Monica~A Lutgendorf and Katie~A Stoll.
\newblock Why 99\% may not be as good as you think it is: limitations of
  screening for rare diseases, 2016.

\bibitem{simon1990sensitivity}
David Simon and John~R Boring~III.
\newblock Sensitivity, specificity, and predictive value.
\newblock In {\em Clinical Methods: The History, Physical, and Laboratory
  Examinations. 3rd edition}. Butterworths, 1990.

\bibitem{balayla2020prevalence}
Jacques Balayla.
\newblock Prevalence threshold and the geometry of screening curves.
\newblock {\em arXiv preprint arXiv:2006.00398}, 2020.

\bibitem{moons1997limitations}
Karel~GM Moons, Gerrit-Anne van Es, Jaap~W Deckers, J~Dik~F Habbema, and
  Diederick~E Grobbee.
\newblock Limitations of sensitivity, specificity, likelihood ratio, and bayes'
  theorem in assessing diagnostic probabilities: a clinical example.
\newblock {\em Epidemiology}, pages 12--17, 1997.

\bibitem{woloshin2020false}
Steven Woloshin, Neeraj Patel, and Aaron~S Kesselheim.
\newblock False negative tests for sars-cov-2 infection—challenges and
  implications.
\newblock {\em New England Journal of Medicine}, 2020.

\bibitem{mcgee2002simplifying}
Steven McGee.
\newblock Simplifying likelihood ratios.
\newblock {\em Journal of general internal medicine}, 17(8):647--650, 2002.

\bibitem{grimes2002uses}
David~A Grimes and Kenneth~F Schulz.
\newblock Uses and abuses of screening tests.
\newblock {\em The Lancet}, 359(9309):881--884, 2002.

\bibitem{weisstein2002ceiling}
Eric~W Weisstein.
\newblock Ceiling function.
\newblock {\em https://mathworld. wolfram. com/}, 2002.

\bibitem{balayla2020derivation}
Jacques Balayla.
\newblock Derivation of generalized equations for the predictive value of
  sequential screening tests.
\newblock {\em arXiv preprint arXiv:2007.13046}, 2020.

\end{thebibliography}

\end{document}